\documentclass[twocolumn,showpacs,floatfix,preprintnumbers,prb]{revtex4-1}
\usepackage{graphicx}
\usepackage{float}
\usepackage{dcolumn}
\usepackage{bm}
\usepackage{amssymb}

\begin{document}

\title{Effects of electron coupling to intra- and inter-molecular vibrational modes on the transport properties of single crystal organic semiconductors}

\author{ C.A. Perroni, V. Marigliano Ramaglia, and V. Cataudella }
\affiliation{CNR-SPIN and Dipartimento di Scienze Fisiche, Univ. di Napoli ``Federico
II'', I-80126 Italy}

\begin{abstract}
Electron coupling to intra- and inter-molecular vibrational modes is investigated in models appropriate to single crystal organic semiconductors, such as oligoacenes. Focus is on spectral and transport properties of these systems beyond perturbative approaches. The interplay between different couplings strongly affects the temperature band renormalization that is the result of a subtle equilibrium between opposite tendencies: band narrowing due to interaction with local modes, band widening due to electron coupling to non local modes. The model provides an accurate description of the mobility as function of temperature: indeed, it has the correct order of magnitude, at low temperatures, it scales as a power-law $T^{-\delta}$ with the exponent $\delta$ larger than unity, and, at high temperatures, shows an hopping behavior with a small activation energy.
\end{abstract}
\maketitle

\section {Introduction}
In the last years, the availability of single crystal organic field-effect (OFET) transistors has represented a step forward in the field of plastic electronics since these systems show a charge mobility at least one order of magnitude larger than that of thin films. \cite{hasegawa} Among them, the most promising and studied are those based on oligoacenes, such as pentacene and rubrene. \cite{morpurgo}

Although the transport properties in the linear regime have been carefully extracted by studying clean single crystal OFET devices, the intrinsic transport mechanism acting in organic semiconductors (OS) is, up to now, not fully understood. Typically, in these materials, the induced doping is not very high (typically much smaller than one charge carrier for ten molecules), so that the mobility is analyzed in transport measurements.  At temperatures close or higher than $100 K$, the mobility $\mu_p$ of these systems exhibits a power-law behavior ($\mu_p \sim T^{-\delta}$, with $\delta \simeq 2$), \cite{morpurgo,ostroverkhova,nature} which cannot be simply ascribed to band transport. \cite{cheng}  Moreover, in some systems, starting from room temperature, there is a crossover from band-like to activated hopping behavior. \cite{corop} For example, in naphthalene and anthracene, along the $a$ and $b$ axis, there is a change in the temperature behavior of mobility, and, along the $c$ axis, an activated regime takes place with a small activated gap of the order of $20 meV$. \cite{warta} Moreover, in pentacene, above room temperature, the mobility shows an upturn with increasing temperature. \cite{morpurgo}

Recently, many photoemission experiments have been performed in pentacene and rubrene. \cite{arpes,arpes1,arpes2,kakuta} The quasi-particle energy dispersion does not exhibit a strong mass renormalization indicating moderate values of electron-phonon (el-ph) coupling. For pentacene, the band structure is reduced only by about $15 \%$ going from $75 K$ to $300 K$.

Typically, the crossover from tunneling to hopping behavior observed in transport properties is ascribed to the formation of local polarons, \cite{cheng1,hannewald} and described with the simplest model for molecular crystals, the Holstein model. \cite{holstein1,holstein2}
The modeling assumes that the most relevant carrier interaction is with local modes whose frequency is lower than the typical hopping energy. \cite{cheng1,hannewald} The used approaches only generalize the previous treatment of Holstein works, \cite{holstein1,holstein2}  starting from the questionable limit of narrow electronic band even for the coupling to low-frequency modes. Actually, recent results, based on the adiabatic approach for the vibrational modes, have pointed out that, up to room temperature, within a parameter range valid for many oligoacenes, polaronic effects do not play a prominent role if low frequency modes are excited in the Holstein model. \cite{meholstein}

Recent {\it ab-initio} calculations have evidenced that charge carriers are mostly coupled to intra-molecular modes with high frequency. Moreover, it has been found that another very relevant interaction is with inter-molecular modes with low frequency in comparison with typical electron hoppings. \cite{corop,bredas} Another important result is that the reorganization energy (related to the polaron binding energy) decreases with increasing the number of benzene rings in oligoacenes (for example, going from naphthalene to pentacene).

In view of {\it ab-initio} results, Holstein-Peierls models have been considered for the description of OS,  \cite{hannewald1,wang,ester} modelling the carrier interaction with low-frequency inter-molecular modes via a modulation of the transfer hopping. Three dimensional models have been mainly applied to naphtalene. \cite{hannewald1} One-dimensional models similar to Su-Schrieffer-Heeger (SSH) model \cite{SSH} have been introduced to study the OS along the relevant stack direction. \cite{troisi_prl,fc,meSSH,vittoriocheck} These models  treat the inter-molecular modes as classical exploiting their low energy in comparison with the electronic hopping scale. The approaches are non perturbative in the electron-phonon coupling. The mobility is quite well described at low temperatures showing a robust power-law $T^{-\delta}$. However, the power-law behavior extends up to high $T$ and there is no sign of upturn in the mobility as function of temperature.

Aim of this work is to clarify the effects of electron coupling to intra- and inter-molecular vibrational modes on the transport properties of prototype single crystal OS, such as oligoacenes. To this aim, a one-dimensional model with coupling to intra- and inter-molecular modes has been analyzed. We will show that the interplay between local and non local el-ph interactions is able to provide a very accurate description of the mobility and to shed light on the intricate mechanism of band narrowing with increasing temperature.

\section {Model}
We consider a one-dimensional model with coupling to intra- and inter-molecular modes similar to one recently introduced, where the treatment only concerns the study of spectral properties. \cite{ciuchi}
The coupling to intra-molecular modes is Holstein-like, that to inter-molecular modes is SSH-like.
It can be summarized in the following hamiltonian:
\begin{equation}
H= H_{el}^{(0)}+H_{Intra}^{(0)}+H_{Inter}^{(0)}+H_{el-Intra}+H_{el-Inter}.
\label{h}
\end{equation}

In Eq. (\ref{h}), the free electronic part $H_{el}^{(0)}$ is
\begin{equation}
H_{el}^{(0)}=-t \sum_{i}  \left( c_{i}^{\dagger}c_{i+1}+ c_{i+1}^{\dagger}c_{i}   \right),
\label{hel}
\end{equation}
where $t$ is the bare electron hopping between the nearest neighbors on the chain, $c_{i}^{\dagger}$ and
$c_{i}$ are the charge carrier creation and annihilation operators, respectively, relative to the site $i$ of a chain with lattice parameter $a$. 
For the transfer hopping the {\it ab-initio} estimate is: $t \simeq 50-100 meV$. \cite{corop} We consider a single-band one-dimensional electronic structure since it represents the simplest effective model in the anisotropic OS to analyze the low energy features responsible for the mobility properties.

In Eq. (\ref{h}), $H_{\alpha}^{(0)}$, with $\alpha= Intra, Inter$ is the Hamiltonian of the free optical molecular modes
\begin{equation}
H_{\alpha}^{(0)}= \sum_{i} \frac{{r}^2_{\alpha,i}}{2 m_{\alpha}}+ \sum_{i} \frac{ k_{\alpha} z_{\alpha,i}^{2}}{2},
\label{hintra}
\end{equation}
where $z_{\alpha,i}$ and ${r}_{\alpha,i}$ are the oscillator displacement and momentum of the mode $\alpha$, respectively, $m_{\alpha}$ the oscillator mass and $k_{\alpha}$ the elastic constant of the mode $\alpha$. For non-local modes, we fix
$y_i=z_{Inter,i}$; for local modes, $x_i=z_{Intra,i}$ and $p_i=r_{Intra,i}$.  The inter-molecular modes are characterized by small frequencies ($\hbar \omega_{Inter} \simeq 5-10 meV$) in comparison with the transfer hopping. \cite{corop,troisi_prl} On the contrary, the most coupled intra-molecular modes have large frequencies ($\hbar \omega_{Intra} \simeq 130-180 meV$). \cite{corop}

In Eq. (\ref{h}), $H_{el-Intra}$ is the Holstein-like Hamiltonian describing the electron coupling to
intra-molecular modes
\begin{equation}
H_{el-Intra}= \alpha_{Intra} \sum_{i} x_i n_i,
\label{hcoupling}
\end{equation}
with $\alpha_{Intra}$ coupling constant to local modes and $n_i=c_{i}^{\dagger}c_{i}$ local density operator. The dimensionless constant
$g_{Intra}= \alpha_{Intra} / \sqrt{2 \hbar m_{Intra} \omega_{Intra}^3}$ is used to describe this el-ph coupling. \cite{holstein1} In single crystal
OS, $g_{Intra}$ is in the weak to intermediate regime (of the order of unity).

Finally, in Eq. (\ref{h}),  $H_{el-Inter}$ represents the SSH-like term with electron coupling to inter-molecular modes
\begin{equation}
H_{el-Inter}= \alpha_{Inter} \sum_{i} (y_{i+1}-y_i) \left( c_{i}^{\dagger}c_{i+1}+ c_{i+1}^{\dagger}c_{i}   \right),
\label{hcoupling1}
\end{equation}
with $\alpha_{Inter}$ coupling constant to non local modes. In the adiabatic regime for non local modes ($\hbar \omega_{Inter} \ll t$), the dimensionless quantity $\lambda_{Inter}=\alpha_{Inter}^2/4 k_{Inter} t $ fully provides the strength of the electron coupling to inter-molecular modes. The typical values of $\lambda$ are in the intermediate coupling regime (of the order of or less than $0.1$).

In the following, we will use units such that lattice parameter $a=1$, Planck constant $\hbar=1$, Boltzmann constant $k_B=1$, and electron charge $e=1$.
We will analyze systems with chain size $L=32$ (where the system reaches the thermodynamic limit) and we will measure energies in units of
$t \simeq 80 meV$.  We fix $\omega_{Intra}=2.0$ as the highest energy scale. \cite{corop} For the calculation of dynamic quantities, an additional small broadening $\Gamma$ is introduced in order to simulate the effect of a tiny disorder.  We fix $\Gamma=0.05 t$, therefore it is the smallest energy scale. The results at low temperatures are only slightly dependent on the value of $\Gamma$.

\section {Method}
Since a very low carrier density is injected into OS, we will study the case of a single particle.

The temperature range where intrinsic effects are relevant is $\omega_{Inter} \leq T \ll t < \omega_{Intra} $. Therefore, the dynamics of intermolecular modes is assumed classical. Actually, the electron dynamics is strongly influenced by the statistical "off diagonal" disorder, that, in the limit of low carrier density, is described by the probability function $P \left( \{ y_i \}  \right) $ of free classical harmonic oscillators:
\begin{equation}
P \left( \{ y_i \}  \right)= \left( \frac{\beta k_{Inter}} {2 \pi} \right)^{L/2}
\exp\left[-\beta\frac{k_{Inter}}{2}\sum_{i}y_{i}^{2}\right],
\end{equation}
with $\beta=1/T$.

At a fixed configuration of non local displacements $\{ y_i \}$, Eq. (\ref{h})  is equivalent to a Holstein model, where the electron hopping between nearest neighbor sites is not homogeneous and depend on the specific pair (see Eq.(\ref{hcoupling1})).  The resulting inhomogeneous Holstein model can be accurately studied within the modified Lang-Firsov approach via a unitary transformation $U$, provided that the system is in the anti-adiabatic regime ($\omega_{Intra} > t $), where the quantum nature of phonons cannot be neglected. \cite{lang,memanganiti} The electron is renormalized by the coupling with local modes increasing its mass, and the oscillators are displaced from their equilibrium position to a distance proportional to the el-ph interaction. In our case, the new Hamiltonian $\tilde{H}=U^{-1} H U$ is built through the variational transformation:
\begin{equation}
U \left( \{ y_j \}  \right)= \exp{ \left[ i g_{Intra} \sqrt{\frac{2}{m_{Intra} \omega_{Intra}} } \sum_i f_i \left( \{ y_j \} \right) p_i n_i \right] } ,
\end{equation}
where the variational parameters $f_i \left( \{ y_j \} \right)$, giving the new centers for the local oscillators at the sites $i$, have to be determined for each fixed configuration of non local displacements $\{ y_j \}$. Within our approach, \cite{memanganiti,mahan} the solution of the full problem can be obtained by minimizing the free energy of the following inhomogeneous effective hamiltonian:
\begin{eqnarray}
&& \tilde{H}_{eff} \left( \{ y_i \}  \right)= H_{Intra}^{(0)} + \sum_{i} \left[ f_i^2 \left( \{ y_j \} \right) -2 f_i \left( \{ y_j \} \right) \right]n_i  + \nonumber\\
&& \sum_{i} \left[ -t + \alpha_{Inter} (y_{i+1}-y_i)  \right] e^{- T_{i,i+1} \left( \{ y_j \} \right)}  \left( c_{i}^{\dagger}c_{i+1}+ h.c.   \right), \nonumber
\end{eqnarray}
where
\begin{equation}
T_{i,j}\left( \{ y_j \} \right)= \frac{ g_{Intra}^2 \left[ f_i^2\left( \{ y_j \} \right) + f_j^2 \left( \{ y_j \} \right) \right](2N_0+1)}{2}
\end{equation}
represents the term giving the reduction of the bond hopping, with $N_0=1/(e^{\beta \omega_{Intra}}-1)$ Bose distribution of the local modes.
The distribution of $f_i$ values depends on the distribution $P \left( \{ y_i \} \right)$. For example, close to $T=0.5 t$, the deviations of  $f_i$ from the average value can become of the order of twenty per cent in the parameter regime investigated in this work. For strong el-ph coupling and high temperatures, the values of $f_i$ are close to one ($T_{i,i+1}$ large), meaning that the system is characterized by small polaron behavior and an incoherent hopping dynamics. For intermediate el-ph coupling and low temperatures, the distribution of  $f_i$ can have an average smaller than unity. This implies that there is a possibility of tunneling dynamics with a renormalized bandwidth ($T_{i,i+1}$ not large).

At fixed configuration $ \{ y_j \}$, one has to diagonalize $\tilde{H}_{eff}$ yielding $L$ eigenvalues $E_n$ and eigenvector components $b_j(n)$, with $j=1,..,L$. In this paper, we will focus on the density of states $D(\omega)\left( \{ y_j \} \right)$:
\begin{equation}
D(\omega)\left( \{ y_j \} \right)=\frac{1}{L} \sum_{i} A_{i,i}(\omega)\left( \{ y_j \} \right),
\end{equation}
where $A_{i,i}(\omega)\left( \{ y_j \} \right)$ represents the diagonal term of the spectral function
\begin{eqnarray}
&& A_{i,i}(\omega) \left( \{ y_j \}  \right)=
\nonumber\\
&& \sum_n | b_i(n)|^2 e^{-T_{i,i}\left( \{ y_j \} \right)} \sum_{- \infty}^{\infty} e^{\beta l \omega_0} I_l
\left( V_i \left( \{ y_j \} \right) \right) B_n(\omega-l \omega_0),
\nonumber
\end{eqnarray}
with $V_i \left( \{ y_j \} \right)=  2 g_{Intra}^2 f_i^2 \left( \{ y_j \} \right) \sqrt{N_0(N_0+1)}$, $I_l(z)$ modified Bessel function of order $l$ and the function $B_n(z)$
\begin{equation}
B_n(z)=\frac{1}{\pi} \frac{\Gamma}{(z-E_n)^2+\Gamma^2}.
\end{equation}
The sum over $l$ provides the contribution of phonon replicas to the density of states. \cite{mahan}

Another central quantity of this work is the mobility as function of the temperature. At fixed configuration $\{ y_j \}$, the mobility
$\mu \left( \{ y_j \} \right)$ of the inhomogeneous Holstein model is determined starting from the real part of the conductivity
$Re[\sigma(\omega)] \left( \{ y_j \} \right)$, taking the limit of zero frequency and dividing for the single particle density $1/L$ of the system. In the linear regime, the real part of the conductivity is derived from the Kubo formula \cite{mahan}
\begin{equation}
Re[\sigma(\omega)] \left( \{ y_j \} \right)=\frac{ \left( 1-e^{-\beta \omega} \right) }{2 \omega} \frac{1}{L} \int_{-\infty}^{\infty} d t e^{i \omega t}
\langle j^{\dagger}(t) j(0) \rangle,
\label{kubo}
\end{equation}
where
\begin{equation}
j=-i \sum_{j,\delta} \delta c^{\dagger}_{j+\delta} c_j h_{j,\delta} \left( \{ y_j \} \right)
\end{equation}
is the current operator, $ h_{j,\delta} \left( \{ y_j \} \right)= t-\alpha \delta (y_{j+\delta}-y_j) $ denotes the generalized hopping and $\delta=-1,1$ indicates the nearest neighbors. In our approach, the current-current correlation function becomes
\begin{equation}
\langle j^{\dagger}(t) j(0) \rangle=\sum_{j,j_1} \sum_{\delta,\delta_1} ( \delta \cdot \delta_1)  h_{j,\delta}  h_{j_1,\delta_1} C_{j,\delta;j_1,\delta_1}(t)
E_{j,\delta;j_1,\delta_1}(t),
\end{equation}
where $C_{j,\delta;j_1,\delta_1}(t)$ is the electron correlation function
\begin{equation}
C_{j,\delta;j_1,\delta_1}(t)=\langle c^{\dagger}_j (t)  c_{j+\delta} (t) c^{\dagger}_{j_1 + \delta_1} (0) c_{j_1} (0)  \rangle_{\tilde{H}_{eff}},
\end{equation}
and $E_{j,\delta;j_1,\delta_1}(t)$ is the phonon correlation function
\begin{equation}
E_{j,\delta;j_1,\delta_1}(t)=\langle X^{\dagger}_j (t)  X_{j+\delta} (t) X^{\dagger}_{j_1 + \delta_1} (0) X_{j_1} (0)  \rangle_{\tilde{H}_{eff}},
\end{equation}
with $X_{i} $ the multi-phonon operator
\begin{equation}
X_{i} = \exp{ \left[ i g_{Intra} \sqrt{\frac{2}{m_{Intra} \omega_{Intra}} } f_i \left( \{ y_j \} \right) p_i \right] }.
\end{equation}
The function $E_{j,\delta;j_1,\delta_1}(t)$ can be written as
\begin{eqnarray}
E_{j,\delta;j_1,\delta_1}(t)&&=e^{- T_{j,j+\delta \left( \{ y_j \} \right)}} e^{- T_{j_1,j_1+\delta_1 \left( \{ y_j \} \right)}}+
\nonumber\\
&&
\left[ E_{j,\delta;j_1,\delta_1}(t) - e^{- T_{j,j+\delta \left( \{ y_j \} \right)}} e^{- T_{j_1,j_1+\delta_1 \left( \{ y_j \} \right)}}  \right].
\nonumber
\end{eqnarray}
With the first term of the previous equation, one can build up the {\it coherent} part of conductivity, where the charge transfer is not accompanied by processes changing the number of phonons. With the second term, retaining only the main autocorrelation contribution $j=j_1$, one gets the {\it incoherent} part of the conductivity, which takes into account inelastic scattering processes of emission and absorption of phonons. \cite{perroni}

If an observable $O$ depends on the displacements $\{ y_i \}$, first one makes the average over the eigenstates and eigenvectors of $\tilde{H}_{eff}\left( \{ y_i \}  \right)$, then over the distribution $P \left( \{ y_i \} \right)$ making the integral
\begin{equation}
\left\langle O \right\rangle=\int  \left( \prod_{i} d y_{i} \right)    P \left( \{ y_i \} \right)
\tilde{O} \left( \{ y_i \} \right)
\label{distri}
\end{equation}
by means of a Monte-Carlo procedure. Actually, we generate a sequence of random numbers distributed according to $P \left( \{ y_i \}  \right)$. For the systems investigated in this paper, a few thousands of iterations are sufficient to get a good accuracy even for dynamic quantities.

The method exposed above is very accurate in the regime $\omega_{Inter} \ll t$ and $\omega_{Intra} > t $ appropriate to OS. It properly takes into account the quantum effects of high frequency local vibrational modes. Moreover, the approach is able to include spatial correlations relevant in quasi one-dimensional systems, in particular vertex corrections in the calculation of mobility.

\section {Results}

\begin{figure}[htb]
\flushleft
\includegraphics[width=0.55\textwidth,angle=0]{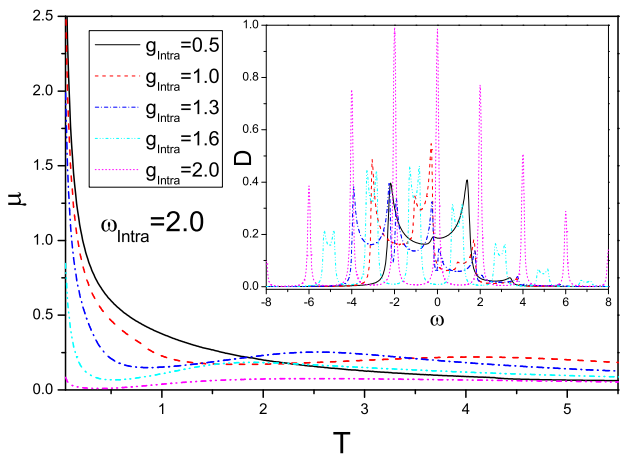}
\caption{Mobility $\mu$ as function of temperature for different values of local coupling $g_{Intra}$.
In the inset, the density of states $D$ as a function of the frequency at $T=0$ for different values of local coupling $g_{Intra}$.}
\label{spectral1}
\end{figure}

First, we analyze the case $\lambda_{Inter}=0$. In Fig. 1, we report the mobility with varying the values of $g_{Intra}$. For weak $g_{Intra}$, a coherent behavior is present with a power-law of the order of $1/T$ at low temperature. As expected, with increasing  $g_{Intra}$, there is a crossover from tunneling to hopping behavior due to the formation of localized small polarons. That occurs at temperatures slightly larger than $0.5 t$ for $g_{Intra}=1.3$. For large values of $g_{Intra}$, the low temperature power-law part is strongly reduced and the activated behavior is followed by the residual scattering regime at very high $T$. \cite{mahan}

In the variational approach, the crossover from band-like to hopping regime is marked by the value $f=1$ on all the sites. However, in the intermediate regime
(for example $g_{Intra}=1.3$), at temperatures characterized by a metallic behavior, the calculation provides values of $f$ smaller than unity ($f$ of the order of $0.6$) suggesting that the tunneling regime is reduced but still present. Moreover, at higher temperatures, where $f=1$ is recovered, the activated gap $\Delta$ in the hopping mobility is not very large: $\Delta \simeq 0.54 t$.

In the inset of Fig. 1, we report the density of states at $T=0$ for the same values of $g_{Intra}$ of the main figure.
As expected, with increasing  $g_{Intra}$, there is a reduction of the fundamental polaronic band with a transfer of spectral weight to high satellite bands appearing at multiples of the vibrational frequency $\omega_{Intra}$. \cite{mahan}

\begin{figure}[htb]
\centering
\includegraphics[width=0.55\textwidth,angle=0]{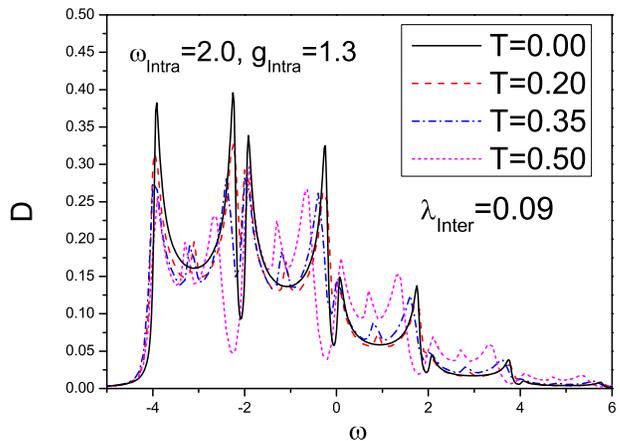}
\caption{The density of states as function of the frequency for different temperatures at $\lambda_{Inter}= 0.09$ and $g_{Intra}=1.3$.}
\label{spectral2}
\end{figure}

The description of mobility based only on the intra-molecular mode electron coupling fails for two reasons in the comparison with low temperature experiments. First, it does not provide large values of the mobility (expressed in our natural units, that is in terms of $\mu_0 = e a^2 / \hbar
\simeq 7 cm^2 /(V \cdot s) $, taking $a=7 \AA$ \cite{corop}). For example, at $T=0.31 t$ (room temperature), $\mu$ is less than $0.5 \mu_0$ for $g_{Intra}=1.3$. Then, the low temperature mobility due local coupling scales as $1/T$.

The next step is to combine the effects of high frequency local vibrational modes with non local low frequency ones. These non local modes are considered as relevant for the description of the mobility up to room temperature. Actually, for rubrene, the carrier mobility is dominated by inter-molecular phonons (the non local coupling is estimated to be in the intermediate regime $\lambda_{Inter} \simeq 0.09$) since the interaction with the intra-molecular modes is almost negligible ($g_{Intra} \simeq 0.7 $). \cite{troisi1} In pentacene, the coupling with local modes is of the same order of that in rubrene. \cite{corop} For these materials, our approach, being the local el-ph coupling perturbative, predicts a slight reduction of mobility in agreement with results present in the literature. \cite{troisi1} On the other hand, going from pentacene to naphthalene, the reorganization energy nearly increases twice suggesting a much stronger coupling with local modes. \cite{corop} Therefore, for systems with reduced number of benzene rings like naphthalene, we expect a larger interplay between intra- and inter-molecular modes within the intermediate el-ph coupling regime for both modes.

In Fig. 2, we report the density of states $D$ for $g_{Intra}=1.3$ and $\lambda_{Inter}=0.09$ at different temperatures.
At $T=0$, where the effect of non local modes is weak, the lowest band is strongly renormalized at lower energies and becomes narrower (about half of the bare band). The intrisic reduction of the bare band provides a simple and direct explanation of the difference in the bandwidth evidenced in the series of oligoacenes from naphthalene (effective band of the order of $40 meV$) to pentacene (effective band of the order of $80 meV$). Indeed, it can be ascribed to the decrease of the reorganization energy that in turn reduces the renormalization effects. \cite{corop}

\begin{figure}[htb]
\centering
\includegraphics[width=0.55\textwidth,angle=0]{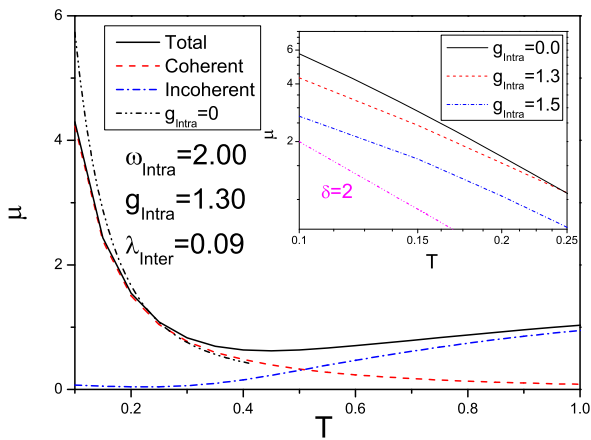}
\caption{Mobility and its different contributions as function of the temperature. In the inset, the mobility as function of temperature for different
values of $g_{Intra}$. For comparison, the behavior with $\delta=2$ is shown.}
\label{density1}
\end{figure}

At finite temperature, the shape of the spectra is changed due to the non local coupling. Actually, any band shows a new small maximum due to the coupling to inter-molecular modes. Moreover, the band narrowing is strongly reduced in the presence of non local coupling. In order to clarify this point, we notice that, for $\lambda_{Inter}=0$, the principal band at $T=0.35$ is reduced of about $42 \%$ of the band at $T=0$. On the other hand, for  $\lambda_{Inter}=0.09$, the principal band at $T=0.35$ is reduced of only $7 \%$ of the band at $T=0$. The narrowing of the principal band results from a subtle equilibrium between the two opposite tendencies. Actually, the coupling to non local modes has the main effect to induce scattering into the single bands of the density of states, preventing the narrowing induced by the coupling to local modes.
The interplay between local and non local modes is able to produce a modest narrowing as function of the temperature even if the coupling to local modes is not
weak. Our prediction is that this effect should be present not only in pentacene, \cite{arpes2} but also in naphthalene and anthracene.

Finally, we analyze the mobility in the intermediate regime (see Fig. 3). As shown in the previous section, the mobility can be divided into two contributions \cite{perroni}: the coherent, where the scattering of the renormalized electron (the only effect due local el-ph coupling is here the reduction of the bandwidth)  with non local modes is included, and the incoherent one, where, in addition to non local modes, scattering with multiple real local phonons is considered. The first term, relevant at low temperatures, bears a strong resemblance with the mobility of the system at $g_{Intra}=0$, even if, as expected, it is smaller. The incoherent term starts at a temperature of about $T=0.25$ and becomes predominant at higher temperatures. Actually, the coherent term only decreases with temperature even if the electron dynamics becomes incoherent. The role of the local coupling is to promote an activated behavior in the incoherent regime. The cooperative effect is able to provide an activation energy $\Delta$ of about $0.25 t$, therefore less than one half of that for $\lambda_{Inter}=0$ and close to that extracted by the data in naphthalene. \cite{warta} Moreover, the upturn in the mobility starts at lower temperatures in comparison with the case $\lambda_{Inter}=0$.

The local coupling is able to affect but to not destroy the low temperature behavior dominated by the non local coupling. In the inset of Fig. 3, we report the low temperature mobility in logarithmic scale. Actually, for $g_{Intra}=0$, the mobility scales as $1/T^{1.89}$, while, with increasing $g_{Intra}$, the power-law becomes slightly less pronounced. In the case $g_{Intra}=1.3$, the mobility goes as $1/T^{1.60}$, still compatible with experiments in naphthalene.
\cite{warta}

In summary, the proposed model is able to capture many features of the mobility in oligoacenes. First of all, it has the correct order of magnitude: for $T=200 K$ ($T \simeq 0.2t$), $\mu \simeq 1.5 \mu_0 \simeq 10 cm^2 /(V \cdot s) $, while at $T=100 K$ ($T \simeq 0.1t$), $\mu \simeq 4 \mu_0 \simeq 28 cm^2 /(V \cdot s) $. These values compare favorably with mobility measurements in naphthalene: for $T=200 K$, $\mu \simeq 4 cm^2 /(V \cdot s) $ in the hole channel and $\mu \simeq 1 cm^2 /(V \cdot s) $ in the electron channel, while at $T=100 K$, $\mu \simeq 20 cm^2 /(V \cdot s) $ in the hole channel and  $\mu \simeq 3 cm^2 /(V \cdot s) $ in the electron channel. \cite{warta} At low temperatures, the mobility scales as a power-law $T^{-\delta}$ with the exponent $\delta \simeq 1.6$ larger than unity even if an intermediate el-ph coupling to local modes is present. Starting from room temperature, the incoherent contribution of mobility begins increasing. At high temperatures, the mobility shows an hopping behavior with a small activation energy. The theory is intended to describe the mobility in the direction where it has the maximum value. Typically, this corresponds to a direction into the growth plane of the OFET on the substrate. \cite{morpurgo} The study of mobility in the other in-plane direction and out of plane is out of the aim of this study.

\section {Conclusions}
The interplay between local and non local el-ph couplings strongly affects the band narrowing and the mobility as function of the temperature. The band narrowing has a polaronic origin but it is strongly dependent on the non local el-ph coupling. In the intermediate coupling regime of the electron with both intra- and inter-molecular modes, the mobility scales as a power-law $T^{-\delta}$ at low temperatures, and it is characterized by hopping behavior with a small activation energy at high temperatures. We believe that the main conclusions will not be qualitatively modified by the inclusion of more realistic lattice structures and interactions since the results rely on general features of different types of el-ph coupling.
The next step will be to investigate the effect of traps on the low temperature mobility generalizing the model presented in this work.

\end{document}